\documentclass[conference]{IEEEtran}
\usepackage[dvips]{graphicx}

\begin{document}
\title{On the Evolution of Programming Languages}

\author{\IEEEauthorblockN{K. R. Chowdhary, Professor}
\IEEEauthorblockA{Dept. of Computer Science  and Engineering\\
Jodhpur Institute of Engineering and Technology, Jodhpur \\
Email: kr.chowdhary@iitj.ac.in}}

\maketitle

\begin{abstract}
%\boldmath
This paper attempts to connects the evolution of computer languages with the evolution of life, where the later has been dictated by \emph{theory of evolution of species}, and tries to give supportive evidence that the new languages are more robust than the previous, carry-over the mixed features of older languages, such that strong features gets added into them and weak features of older languages gets removed. In addition, an analysis of most prominent programming languages is presented, emphasizing on how the features of existing languages have influenced the development of new programming languages. At the end, it suggests a set of experimental languages, which may rule the world of programming languages in the time of new multi-core architectures.


\end{abstract}
 
\IEEEpeerreviewmaketitle

\section{Introduction}
   
A fundamental difference between the human and computer languages is that computer languages are designed and implemented by a team of few persons, where the concept originator in most cases is single person. However, in human languages, the language development is evolutionary, it takes very long time to develop, new words keep on adding, old gets eliminated over a time due to many reasons, the most important one of that is that the respective concept gets disappeared. The arrival of news ones are  due their prevalence, convenience, and strengths in expression. Many such changes have taken place in human languages over the time in their evolution process. All these evolutionary criteria are not taken and however the two patterns still holds true in case of computer languages.

The programming languages  have existed for over 200 years, since the invention of the punch-card-programmable {\em Jacquard loom}. It was not a programming language in the modern sense -- there was no computation and no logic -- but it started a cascade that would eventually lead to Charles {\em Babbage’s Analytical Engine}, and Ada Lovelace’s 1842 construction of his work which led to the first computer program.

Machine-specific assembly language in the 1940s was probably the first human-readable programming language, but by the 1950s computer engineers realized that assembly language was far too laborious and error-prone to build entire systems out of it, and thus in 1955 the first modern programming language was born: FORTRAN (FORmula TRANslation). The COBOL (COmmon Business Oriented Language), LISP (LISt Processing language), ALGOL (ALGOrithmic Language) followed in the next few years, it is said that the rest is history. Almost every language today is a derived from one of these first four languages, and indeed, FORTRAN, LISP, and COBOL are still actively used by large number of institutions, and many applications of  these are still running and are maintained.  
   
By 1964, BASIC (Beginner's all purpose Symbolic Instruction Code) had been invented, and then C was released in 1969. Unix was famously re-written into C -- the first major Operating System, to not be written in assembly language -- and today, Linux is written almost entirely in C, and both Windows and Mac OS X have large amount of their code written in C.

\section{Evolution of Computer languages}   

The theory of evolution~\cite{darwin} applies to large extent on the evolution of programming languages also. The theory states that new population is generated from old, through the operations of \emph{cross-over, mutation}, and \emph{selection} (of the fittest). But, some populations should exist to start with. Once, few programming languages (say $X, Y, Z$) were created in the 50s and 60s, rest all the languages which came into existence, carried their features as follows:

\begin{itemize}
\item some features from $X$, some from $Y$, and some from $Z$, in different proportions; thus implementing the operation of \emph{cross-over} of evolution theory;

\item in the newly created languages, some features of old languages totally disappeared, and some totally new features got added first time, thus implementing the operation of \emph{mutation}, of evolution theory, and;

\item only good quality of languages' features were considered in the new languages, for example, C carried the control features of Fortran but not the \emph{goto}, it carried the subroutine call in C, but a better version of that - {\em recursion}. And, when new languages came into existence, the old became extinct; thus implementing the \emph{natural selection} operation of evolution theory.  
\end{itemize}

The theory of evolution~\cite{darwin} says that the size of the population remains more or less same. As the new programming languages emerged (e.g., C++, Java), the old slowly died (e.g., Fortran, ALGOL); similarly in the scripting languages, arrival of Perl and Python, made AWK, Shell programming extinct, the database MYSQL made Dbase, foxbase extinct, and so, on. From the table~\ref{progl}, showing the statistics of programming languages, we note that the production rate of new programming languages, approximating to 11 per decade, is stable, over the span of 60 years life of these languages.

This evolution goes on, and will bring new languages for multicore platform, large majority of the previous languages extinct, however, the most features of older languages will be carried over. The table~\ref{hll}  is list of programming languages of significance, which were born at different times, given along with as time stamp of their birth year. And roughly the same amount is becoming obsolete. Thus, the population, remains more or less stable. Hence, it can be hypothesized that, rule of survival of the fittest also holds true in the case programming languages too.

%table I
\begin{table}[!t]
\caption{High-Level Languages.}
\label{hll}
\begin{center}
\begin{tabular}{ll}
1951: Regional Assembly Lang. & 1983: Ada \\
1952: Autocode & 1984: Common Lisp \\
1954: IPL (forerunner to LISP) & 1984: MATLAB \\
1955: FLOW-MATIC  & 1985: Eiffel \\
\hspace{0.8cm} (forerunner to COBOL) & 1986: Object-C\\
1957: FORTRAN (First compiler) & 1986: Erlang \\
1957: COMTRAN  & 1987: Perl \\
\hspace{0.8cm}(forerunner to COBOL) & 1988: Tcl\\
1958: LISP & 1988: Mathematica \\
1958: ALGOL 58 & 1989: FL (Backus)\\
1959: FACT  & 1990: Haskell  \\
\hspace{0.8cm}(forerunner to COBOL) & 1991: Python\\
1959: COBOL & 1991: Visual Basic \\
1959: RPG & 1991: HTML \\
          & (Mark-up Language)\\
1962: APL & 1993: Ruby \\
          &  1993: Lua  \\
1962: Simula & 1994:CLOS (part of \\
1962: SNOBOL & ANSI Common Lisp) \\
             & 1995: Java \\
1963: CPL (forerunner to C) & 1995: Delphi \\
1964: BASIC & (Object Pascal)\\
1964: PL/I & 1995: JavaScript \\
1967: OyII (forerunner to C) & 1995: PHP \\
1968: Logo & 1996: WebDNA \\
1969: B (forerunner to C) & 1997: Rebol  \\
1970: Pascal & 1999: D \\
1970: Forth & 2000: ActionScript\\
            & 2001: C\# \\
1972: C & 2001: Visual Basic\\
1972: Smalltalk & 2001: .NET \\
1972: Prolog & 2002: F\# \\
1973: ML & 2003: Groovy \\
1975: Scheme & 2003: Scala  \\
1978: SQL  & 2003: Factor \\
1980: C++ (as C with classes, & 2007: Clojure  \\
\hspace{0.8cm}  name changed in July 1983)& 2009: Go \\
  & 2011: Dart\\
\end{tabular} 
\end{center}
\end{table}
  {}

%table II   
\begin{table}[!t]
\renewcommand{\arraystretch}{1.3}
\begin{center}
\caption{Number of New Programming languages born, in year-bands}
\label{progl}
\begin{tabular}{lll}
\hline
\textbf{S.No.} & \textbf{Year} & \emph{No. of languages}\\
\hline
1 & 1951-60 & 11\\
2 & 1961-70 & 12\\
3 & 1971-80 & 07\\
4 & 1981-90 & 12\\
5 & 1991-00 & 15\\
6 & 2001-10 & 09\\
\hline
\end{tabular}
\end{center}
\end{table}

\subsection{Languages Classification}

The high level languages (HLL) developed belonged to many different classes, due to their efficiency and features better suiting to specific problem nature. For example,  the complexity and cryptic nature of machine languages contributed to creation of procedure oriented high level languages, like, Fortran and COBOL. C was better suited for system programming. When the programs became larger and unmanageable, there was need to have languages to handle complexity of huge programs. This gave birth to object-oriented languages. The present classification of programming languages is shown in table~\ref{class}.
 
%table3
\begin{table}[!t]
\renewcommand{\arraystretch}{1.3}
\begin{center}
\caption{Classification of Programming languages.}
\label{class}
\begin{tabular}{ll}
\hline
S. No. & Language class\\
\hline
1. & Assembly languages (specific to Processor)\\
2. & Authoring languages (HTML, Tutor)\\
3. & Compiled languages (Fortran, C, Pascal)\\
4. & Command-line interface languages (OS command, shell)\\
5. & Data flow languages (VHDL, Labview)\\
6. & Data oriented Languages (SQL, Foxpro)\\
7. & Embedded languages (Lua, other scripting languages.)\\
8. & Functional languages (Lisp, Haskell, Erlang) \\
9. & Interpreted languages (Java, Python, Ruby, Perl, PHP,\\
   & PostScript, Python, Lisp, Logo)\\
10. & Logic based languages (Prolog)\\
11. & Macro-languages (ML1, M4, MINIMAC, SAM76)\\
12. & Object Oriented languages (C++, Java)\\
13. & Procedural languages (COBOL, Fortran, C)\\
14. & Rule-based languages (Prolog)\\
15. & Scripting languages (python, perl, awk, sed)\\
16. & Stack-based languages (Forth, RPL, PostScript, BibTeX)\\
17. & Synchronous languages (Argos, Atom, Averest, LabVIEW, SIGNAL)\\ 
18. & Syntax handling languages (Ada, Bash, BASIC, C) \\ 
19. & Visual languages (VB .NET, Visual C++)\\
20. & XML based languages (AIML, LGML, LOGML)\\
\hline
\end{tabular}
\end{center}
\end{table}

\subsection{Natural Programming Languages and Environments} 
 
One way to define programming is the process of transforming a mental plan in familiar terms into one compatible with the computer~\cite{bradam}. The closer the language is to the programmer’s original plan, the easier this refinement process will be. This is closely related to the concept of directness that, as part of direct manipulation, is a key principle in making user interfaces (UI) easier to use.  By ``natural,"  means ``faithfully representing nature or life," which here implies it works in the way people expect. Natural programming is aimed for the language and environment to work the way that non-programmers expect. 

Conventional programming languages require the programmer to make tremendous transformations from the intended tasks to the code design. For example, a typical program to add a set of numbers in C uses three kinds of parentheses and three kinds of assignment operators in five lines of code, whereas a single ``SUM" operator is sufficient in a spreadsheet. It is argued that if the computer language were to
enable people to express algorithms and data more like their natural expressions, the mental transformation effort would be reduced.

Similarly, debugging activities could benefit from being more natural. Debugging is described as an exploratory activity aimed at investigating a program's behavior, involving several distinct and interleaving activities:

\begin{itemize}
\item Hypothesizing what run-time actions caused failure;
\item Observing data about a program's run-time state;
\item Restructuring data into different representations;
\item Exploring restructured run-time data;
\item Diagnosing what code caused faulty run-time actions; and
\item Repairing erroneous code to prevent such actions.
\end{itemize}

\section{Implementing Programming languages}

Some of the terms for implementation of languages are:\\

\emph{Processor-} a computer program which processes other computer programs.\medskip

\emph{Compiler-} A processor which converts a program written in a programming language (called the ``source code") into code which a computer can execute (called ``object code").\medskip

\emph{Interpreter-} a processor which accepts a program written in a source code, converts it into some readily executable form, and performs a controlled execution. The readily executable form may or may not be object code.\medskip

 \emph{Translator-} a processor which converts one source code into another.

\subsection{Structures}

Like the natural languages have structures at word level, sentence level, the meaning associated with sentences, the programming languages too have these structures.

\emph{Lexical structures:} This corresponds to world level structures and decides about the token in the language. This is recognized by abstract machines, called finite automata~\cite{krc}.\medskip

\emph{Syntax structures:} This sentence level structures, expressed by given by parse-tree, and recognized by the abstract machine - push-down automata. \medskip

\emph{Semantic structures:} It is given by association of nodes in parse-tree.\medskip

\emph{Ambiguity in languages:} If there is more than one syntax trees for the same sentence, the language as well as the grammar is ambiguous.\medskip

\emph{Semantics:} Semantics based on the lexicons.

\subsection{Compiler Writing}
 
A major and difficult task for implementation of any new programming language is compiler writing. Basically, compilers are organized in the following manner~\cite{glassrl};

\begin{enumerate}
\item The source program to be processed is read into memory as a stream of characters.
\item The character stream is processed in character groups, where a group is a language keyword (IF, GO TO, DO, etc.), a parameter name, a literal (numeric, alphanumeric, etc.) value, an operator, etc. This process is often called "parsing" because of its similarity to the parsing of sentences in a study of natural languages. 

\item As the character groups are analyzed, the information thus gathered is encoded into tables of information (such as a name list) and sequential streams of information (such as generated object code). 

\item Following the processing of the complete source program, the sequential object code stream is analyzed, and missing information is filled in from the appropriate tables.
\end{enumerate}

The Compiler writers employ widely different implementation techniques. However, there
are certain building blocks which are common to many implementations. They may
differ in form, but the functions are generally the same. Among these basic functions are:

\begin{itemize}
\item get next (source text) character
\item read next (source text) element (repeat getchar() above)
\item relate name to name-list
\item obtain statement type
\item process statement type
\item generate intermediate object code
\item read intermediate object code
\item generate or interpret final object code
\item popup/pushdown
\item skip to Specified (source or intermediate) character
\item set character/operator flag
\end{itemize}

\section{Embeddability of new Languages}

Scripting languages are an important element in the current landscape of programming languages. A key feature of a scripting language is its ability to integrate with a system language. This integration takes two main forms: \emph{extending} and \emph{embedding}~\cite{robertoie}. In the first form, you extend the scripting language with libraries and functions written in the system language (e.g., in SQL) and write your main program in the scripting language (e.g., in HTML). In the second form, you embed the scripting language in a host program (written in the system language) so that the host can run scripts and call functions defined in the scripts; the main program is the host program. In this setting, the system language is usually called the host language.

\subsection{Embedding}

At first sight, the embeddability of a scripting language seems to be a feature of the implementation of its interpreter. Given any interpreter, we can attach an \emph{API} (Application Program Interface) to it to allow the host program and the script to interact.  The typical host language for most scripting languages is \emph{C}, and APIs for these languages are therefore mostly composed of functions plus some types and constants. This imposes a natural but narrow restriction on the design of an \emph{API} for a scripting language: it must offer access to language features through this narrow path. Syntactical constructs are particularly difficult to get through. For example, in a scripting language where methods must be written lexically inside their classes, the host language cannot add methods to a class unless the API offers suitable mechanisms.

A key ingredient in the API for an embeddable language is an eval function, which executes a piece of code. In particular, when a scripting language is embedded, all scripts are run by the host calling \texttt{eval}.  An eval function also allows a minimalist approach for designing an API. With an adequate eval function, a host can do practically anything in the script environment: it can assign to variables \texttt{(eval``$a$ = 25")}, query variables \texttt{(eval``return $a$")}, call functions \texttt{(eval``foo(35,'stat')")}, and so on. Data structures such as arrays can be constructed and decomposed by evaluating proper code. For example, again assuming a hypothetical eval function, A $C$ language code shown in below would copy a $C$ array of integers into the script.

\begin{verbatim}
//Passing an array through API with eval.
   
void copy (int ar[], int n) {
int i;
eval(“ar = {}”); // create empty array 
for (i = 0; i < n; i++){
  char buff[100];
  sprintf(buff, “ar[%d] = %d”, i + 1, ar[i]);
  eval(buff); /* assign i-th element */
 }
}
\end{verbatim}

\subsection{Control}

The first problem related to control that every scripting language must decide is ``where to keep the main function?". When we use the scripting language embedded in a host, we want the language to be a library, with the main function in the host. However, for many applications, we want the language as a standalone program with its own main function.

The  scripting language \emph{Lua} solves this problem with the use of a separate standalone program. Lua itself is entirely implemented as a library, with the goal of being embedded in other applications. The lua command-line program is a small application that uses the Lua library as any other host to run pieces of Lua code. The code given below is a bare-bones version of this application.  

\begin{verbatim}
// A Lua application.
#include “lualib.h”
#include <stdio.h>
#include “lauxlib.h”
int main (void) {
char text[80];
lua_State *X = luaL_newstate(); /* create
a new state */  
luaL_openlibs(X); /* open libraries */
/* reads text lines and execute */
while (fgets(text, sizeof(text), stdin)
        != NULL) {
//compile text to a function
  luaL_loadstring(X, text); 
  lua_pcall(X, 0, 0, 0);
 /* call the function */
}
lua_close(X);
return 0;
}
\end{verbatim}

\section{Future of Programming languages}
 
Is there a need for another programming language? There are already abundant languages and no shortage of choices. There are procedural languages, functional languages, object-oriented languages, dynamic languages, compiled  languages, interpreted languages, dynamic languages, and scripting languages, and it is not possible for any developer to even learn, keep apart the development.

But still, new languages emerge with surprising frequency. Some are designed by students or hobbyists as personal projects. Others are the products of large IT vendors. Even small and mid-size companies are getting in on the action, creating languages to serve the needs of their industries. Why there is need to reinvent the existing features?

The answer is that, as powerful and versatile as the current languages may be, no single syntax is ideally suited for every purpose. Over and above, programming itself is constantly evolving. The rise of multi-core CPUs, cloud computing, mobility, and distributed architectures have created new challenges for developers. Adding support for the latest features, paradigms, and patterns to existing languages, especially popular ones, can be prohibitively difficult. Sometimes the best answer is to start from scratch.

Following are some cutting-edge experimental programming languages, each of which approaches the art of software development from a fresh perspective, tackling a specific problem or a unique shortcoming of today's more popular languages.  

\subsection{Go}

Go is a general-purpose programming language suitable for everything from application development to systems programming.  It more like C or C++ than Java or C\#. But like the latter languages, Go includes modern features such as garbage collection, runtime reflection,  support for concurrency, and is meant to be easy to program.  

\subsection{F\#}

Functional programming has long been popular with computer scientists and academia, but pure functional languages like Lisp and Haskell are often considered unworkable for real-world software development. One common complaint is that functional-style code can be difficult to integrate with code and libraries written in imperative languages like C++ and Java.

F\# is a Microsoft language designed to be both functional and practical, it is  first-class language on the .Net Common Language Runtime (CLR), it can access all of the same libraries and features as other CLR languages, such as C\# and Visual Basic. F\# also offers constructs to aid asynchronous I/O, CPU parallelization, and off-loading processing to the GPU. F\# compiler and core library are available under the Apache open source license; you can start working with it for free and even use it on Mac and Linux systems.

\subsection{Dart}

It is an Experimental programming language for web programming, created by google, and its performance improves to better and better as size becomes larger and larger, which is not true for JavaScript. Like JavaScript, Dart uses C-like syntax and keywords. One significant difference, however, is that while JavaScript is a prototype-based language, objects in Dart are defined using classes and interfaces, as in C++ or Java. Dart also allows programmers to optionally declare variables with static types. The idea is that Dart should be as familiar, dynamic, and fluid as JavaScript, yet allow developers to write code that is faster, easier to maintain, and less susceptible to subtle bugs. It's designed to run on either the client or the server both.

\subsection{Opa}

Web development is a complicated task, requiring code in multiple languages: HTML and JavaScript on the client, Java or PHP on the server, SQL in the database, and so on.

Opa does not replace any of these languages individually. Rather, it seeks to eliminate them all at once, by proposing an entirely new paradigm for Web programming. In an Opa application, the client-side UI, server-side logic, and database I/O are all implemented in a single language, Opa.

Opa accomplishes this through a combination of client- and server-side frameworks. The Opa compiler decides whether a given routine should run on the client, server, or both, and it outputs code accordingly.
  
Opa's runtime environment bundles its own Web server and database management system, which cannot be replaced with stand-alone alternatives. That may be a small price to pay, however, for the ability to prototype sophisticated, data-driven Web applications in just a few dozen lines of code. Opa is open source and available now for 64-bit Linux and Mac OS X platforms.

\subsection{Zimbu}

It aims to be a fast, concise, portable, and easy-to-read language that can be used to code anything from a GUI application to an OS kernel. It uses C-like expressions and operators, but its own keywords, data types, and block structures. It supports memory management, threads, and pipes.

Portability is a key concern. Although Zimbu is a compiled language, the Zimbu compiler outputs ANSI C code, allowing binaries to be built only on platforms with a native C compiler.

\subsection{X10}

Parallel processing was once a specialized niche of software development, but with the rise of multi-core CPUs and distributed computing, parallelism is going mainstream. Unfortunately, today's programming languages are not keeping pace with the hardware. To counter it, IBM Research is developing X10, a language designed specifically for modern parallel architectures, with the goal of increasing developer productivity ``times 10."

X10 handles concurrency using the partitioned global address space (PGAS) programming model. Code and data are separated into units and distributed across one or more ``places," making it easy to scale a program from a single-threaded prototype (a single place) to multiple threads running on one or more multi-core processors.

X10 code most resembles Java; in fact, the X10 runtime is available as a native executable and as class files for the JVM. The X10 compiler can output C++ or Java source code. The compiler and runtime are available for various platforms, including Linux, Mac OS X, and Windows.

\subsection{Chapel}

Chapel is Cray's first original programming language, was designed with supercomputing and clustering in mind. Among its goals are abstracting parallel algorithms from the underlying hardware, improving their performance on architectures, and making parallel programs more portable.

Chapel's syntax draws from numerous sources. In addition to the usual suspects (C, C++, Java), it borrows concepts from scientific programming languages such as Fortran and Matlab. its more compelling features is its support for ``multi-resolution programming," which allows developers to prototype applications with highly abstract code and fill in details as the implementation becomes more fully defined.

At present, it can run on Cray supercomputers and various high-performance clusters, but it is portable to most Unix-style systems, Windows with Cygwin.

\subsection{Fantom}

Should you develop your applications for Java or .Net? If you code in Fantom, you can take your pick and even switch platforms midstream. This is because Fantom is designed from the ground up for cross-platform portability. The Fantom project includes not just a compiler that can output bytecode for either the JVM or the .Net CLI, but also a set of APIs that abstract away the Java and .Net APIs, creating an additional portability layer.

While it remains inherently C-like, it is also meant to improve on the languages that inspired it. It tries to strike a middle ground in some of the more contentious syntax debates, such as strong versus dynamic typing, or interfaces versus classes. It adds easy syntax for declaring data structures and serializing objects. And it includes support for functional programming and concurrency built into the language.

\section{Conclusion}

This paper has presented a broad history as well as the evolution of programming languages, and relates the evolution of languages with the evolution of species, given and demonstrated in the Darwin's Theory of evolution. In addition, it has presented the embedding features of scripting languages, and survey of a few experimental languages, which may rule future programming in the multi-core era. 

In pursuit of the searching goal of an ideal programming language, if studies are carried out to examine the language and structure that children and adults naturally use in solving problems before they have been exposed to programming, it can be most natural to conclude a most appropriate language. For this, participants can be presented with programming tasks and asked to solve them on paper using whatever text or diagrams they want.

\end{document}